%% file: main.tex
\documentclass{llncs}

\input{macro}

\begin{document}

\title{SMT Solving for Vesicle Traffic Systems in Cells}

\author{Ashutosh Gupta$^1$ \and Ankit Shukla$^1$ \and
  Mandayam Srivas$^2$ \and Mukund Thattai $^3$}
\institute{TIFR Mumbai \and CMI, Chennai \and NCBS, Bangalore}


\date{\today}

\maketitle

\begin{abstract}
\input{abstract}
\end{abstract}

\section{Introduction}
\label{sec:intro}
\input{intro_MTedit}

\section{Biological Problem}
\label{sec:bio}
\input{bio_MTedit}

\section{The Problem Encoding}
\label{sec:encoding}
\input{encoding}

\section{Implementation and Experiments}
\label{sec:experiments}
\input{experiments}

\section{Related Work}
\label{sec:related}
\input{related}

\section{Conclusion}
\label{sec:conclusion}
\input{conclusion}

\bibliographystyle{unsrt}
\bibliography{biblio}

\end{document}

%% file: macro.tex
\usepackage{mathtools}
\usepackage{makecell}
\usepackage{graphicx}

\usepackage{enumitem}

\usepackage{amsmath,amssymb}

\renewcommand{\figurename}{Fig.{}}

\usepackage{changepage}

\usepackage[utf8x]{inputenc}

\usepackage{textcomp,marvosym}

\usepackage{cite}

\usepackage{nameref,hyperref}

\usepackage[right]{lineno}

\usepackage{microtype}
\DisableLigatures[f]{encoding = *, family = * }

\usepackage[table]{xcolor}

\usepackage{todonotes}

\usepackage{array}

\usepackage{listings} 

\usepackage{subcaption}

\usepackage[british]{babel}
\usepackage{hhline}
\usepackage{multirow}
\usepackage[figurename=Fig]{caption}

\usepackage{tkz-orm}

\usepackage{verbatim}
\usepackage{pgfplots}

 \usepackage{tikz}
 \usepackage{xparse}
\usetikzlibrary{matrix,backgrounds}
\pgfdeclarelayer{myback}
\pgfsetlayers{myback,background,main}

\tikzset{mycolor/.style = {line width=1bp,color=#1}}%
\tikzset{myfillcolor/.style = {draw,fill=#1}}%

\NewDocumentCommand{\highlight}{O{blue!40} m m}{%
\draw[mycolor=#1] (#2.north west)rectangle (#3.south east);
}

\NewDocumentCommand{\fhighlight}{O{blue!40} m m}{%
\draw[myfillcolor=#1] (#2.north west)rectangle (#3.south east);
}
 \usetikzlibrary{matrix,decorations.pathreplacing, calc, positioning}

\newcolumntype{+}{!{\vrule width 2pt}}

\newlength\savedwidth

\newcommand{\booleans}{\mathbb{B}}

\newcommand{\maps}{\rightarrow}

\newcommand{\powerset}[1]{2^{#1}}
\newcommand{\intersection}{{\cap} }

\newcommand{\limplies}{\Rightarrow}

\newcommand{\lequiv}{\Leftrightarrow}

\newcommand{\nodes}{N}
\newcommand{\mols}{M}
\newcommand{\nlabel}{L}
\newcommand{\edges}{E}
\newcommand{\pairs}{P}
\newcommand{\nodef}{f}
\newcommand{\edgef}{g}

\newcommand{\zthree}{\textsc{Z3}}

\newtheorem{df}{Definition}



%% file: abstract.tex
In biology, there are several questions that translate to combinatorial
search. For example, vesicle traffic systems that move cargo within
eukaryotic cells have been proposed to exhibit several graph properties
such as three connectivity. These properties are consequences of underlying
biophysical constraints. A natural question for biologists is: what are the
possible networks for various combinations of those properties? In this
paper, we present novel SMT based encodings of the properties over vesicle traffic
systems and a tool that searches for the networks that satisfies the
properties using SMT solvers. In our experiments, we show that our tool
can search for networks of sizes that are considered to be relevant by
biologists.



%% file: intro_MTedit.tex


The cytoplasm of eukaryotic cells is broken into membrane-bound compartments known as organelles. Cargo is moved between these compartments in small membrane-bound transporters known as vesicles \cite{stenmark2009rab}. This organization resembles a transport logistics network (compartments are nodes, vesicles are directed edges) and is commonly known as the vesicle traffic system (VTS). Vesicle traffic underpins nearly every aspect of eukaryotic cellular physiology, including human cellular physiology. Understanding how this system functions is therefore one of the central challenges of cell biology.

Cell biologists have identified many molecules that drive vesicle traffic~\cite{dacks2007evolution}. The key steps are the creation of vesicles loaded with cargo from source compartments (budding), and the depositing of those vesicles and cargo into target compartments (fusion) \cite{munro2004organelle}. Budding is regulated by proteins known as coats and adaptors that select cargo. Fusion is regulated by proteins known as SNAREs and tethers, that ensure that vesicles fuse to the correct target \cite{mani2016stacking}. Abstractly, we think of SNARE proteins as labels that are collected from the source compartment by vesicles. There is a corresponding set of label molecules on potential target compartments. If the two sets of labels (on vesicles and targets) are compatible, then the vesicle will fuse to the target. Importantly, it is likely that SNARE proteins can be regulated by other molecules, thereby being in active or inactive states. However, SNARE regulation is not well understood.

There has been very little work done on how these molecular processes are integrated across scales to build the entire VTS. We have recently used ideas of graph theory to address this question \cite{mani2016stacking}. We have shown that SNARE regulation places constraints on the structure of the traffic network \cite{shukla}. One informative constraint is graph connectivity. The nature of SNARE regulation determines whether graph topologies of particular levels of connectivity are biologically realizable. For example, if SNAREs are completely unregulated and thus always active, they cannot be used to generate a traffic network. If these SNAREs are regulated by relatively simple rules, then only highly connected traffic networks can be generated using them. In this way, biological inputs (about SNARE regulation) lead to clear testable predictions (about graph topologies).

The analysis of VTSs is a difficult computational problem because of the combinatorial scaling of graph topologies and regulatory rules. Many questions require us to check useful properties over all possible graphs and regulatory rules. Formal verification tools such as model checkers \cite{clarke1996symbolic, biere2003bounded, clarke2008birth, cimatti2000nusmv, holzmann1997model} and SAT solvers \cite{moskewicz2001chaff,een2004extensible} allow us to do this symbolically for graphs of finite size, without enumerating all instances. In a recent work, we employed the model checker CBMC \cite{CKY03, ckl2004} meant to analyze ANSI C programs for studying this problem. VTSs and SNARE regulations were modeled as C code using arrays to represent graphs and Boolean tables. Queries about graph connectivity requirements for different variations of SNARE regulations were modeled as logical assertions to be checked by the model checker. CBMC uses its own built-in encodings to reduce the analysis problem to a Boolean satisfiability problem which is solved using SAT solvers.
However, the scalability of our recent approach was limited (up to
VTSs with 6 nodes) due to the encoding used to model VTSs was not completely
fine-tuned to the structure of the problem.
In particular, we had to encode conditional reachability between
nodes (refer sec. 3, stability condition), and solved in CBMC using a
combination of non-determinism and partial enumeration that is
inefficient.



In this paper, we have developed a novel SMT based encoding for
searching VTSs that satisfy the useful properties.
We have improved the encoding of some of the conditions for VTSs.
Especially, we have recursively defined the reachability condition
such that that we avoid the inefficient enumeration.
%
%
%
We have implemented the encoding using Z3~\cite{z3} python API and
searched for VTS satisfying various properties upto size 14-18 nodes
as compare earlier tool that could only search graph upto
size~6\footnote{The original experiments in~\cite{shukla} scaled upto 8.
However, our current experiments only scale upto 6, because
we are using less powerful machine.} nodes.
Our tool is useful to cell biologists since VTSs of real cells have
approximately 10 compartments.



The following are the contributions of this work:
\begin{itemize}
\item Novel encodings of reachability and 3-4 connectivity
\item Direct encoding into the SMT solver
\item A user friendly and scalable tool based on well known SMT solver Z3
\end{itemize}



%% file: bio_MTedit.tex

In this section, we will describe the basic constraints imposed by cell biology. These are all incorporated into an abstract model of a VTS, whose properties will then be explored using SMT solvers.

\textbf{The cell as a transport graph:} We consider a cell to be a collection of compartments (nodes) and vesicles (edges), thus defining a transport graph. Every compartment or vesicle has a set of molecular labels, such as SNARE proteins, associated with it.

\textbf{Molecular flows and steady state:} Each edge is associated with a flux of all the molecular types carried by the corresponding vesicle. The total amount of each molecular type on each compartment can therefore increase or decrease. We assume the cell is in a steady state where each compartment’s composition does not vary over short time scales. Therefore, incoming and outgoing fluxes are balanced for each molecular type at each compartment; it is the \textit{stability condition}.

\textbf{Vesicle targeting driven by molecular interactions:} Once a vesicle has budded out of the source, the molecules it carries determine its properties. In particular, for any given pair of a vesicle and a compartment, the set of SNARE proteins that label the former and latter influence whether the vesicle will fuse to that compartment. Biophysically, fusion requires a direct physical interaction between at least one SNARE type on the vesicle and one SNARE type on the compartment. SNAREs fall into two classes (known as Q and R in the cell biology literature) and vesicle fusion requires a pairing of a Q-SNARE with an R-SNARE. The list of molecular pairs that can drive a fusion event is given in a pairing matrix between Q and R SNAREs. Without loss of generality we assume equal numbers of Q and R SNARE types.

\textbf{Molecular regulation:} We assume that for fusion to occur, the pair of SNARE types involved on the vesicle and compartment must both be in an active state. Whether these SNAREs are active or inactive depends on the other molecules found on the vesicle or compartment, respectively. We test many different versions of this kind of molecular regulation. Most generally, the activity state of a given SNARE can be a Boolean function of all the molecular types on a compartment or vesicle. We have also tested \cite{shukla} a particularly simple regulation mechanism in which two SNAREs that can pair to drive fusion inhibit one another; this is the \textit{pairing inhibition}. This is motivated by the idea that pairing must generate an inactive bi-molecular complex.

\textbf{Properties of a VTS that satisfies all cell-biological constraints:} Suppose we are given a particular transport graph, a particular labeling of all the compartments and edges, a particular fusion pairing matrix, and a particular regulatory model. This information is then sufficient to check the following properties, which summarise the cell-biological constraints described above:
\begin{enumerate}
\item We can determine which molecules are active on every compartment or vesicle.
\item For every vesicle fusing to a compartment, we can determine whether there exists an active pair (one molecule on the vesicle, one on the compartment) which drives that fusion event.
\item For every vesicle-compartment pair where the vesicle does not fuse to the compartment, we can verify that there is no pairing of active molecules on the
vesicle and compartment that could drive their fusion.
\item We can verify that every molecular type entering a compartment also leaves the compartment, and also that every molecular type entering a set of compartments also leaves that set. This is the steady state condition. In the biological literature this is often referred to as “homeostasis” and is a widespread and well-accepted assumption about cellular behaviour, at least over timescales of minutes to hours \cite{mani2016stacking}.
\end{enumerate}

The biological problem often boils down to such an analysis. A cell biologist might determine which molecules flow between which set of compartments, and biochemical experiments could be used to see how these molecules activate one another. We can then ask: is this a complete and consistent description? That is, do all the required properties listed above hold, given what the experimentalists have told us? It is often the case that biological data is missing. For example, only a few of the dozens of molecules involved in real VTSs have been mapped out. Therefore, it is extremely likely that the description provided by the cell biologist is incomplete. We can use our model to find out which properties have failed to hold, and thus prescribe new experiments in order to fill in the missing information.

Can we find a simple test to see whether any information is missing, given the experimental data? We have shown that graph k-connectedness furnishes precisely such a test \cite{shukla}. If the data provided by cell-biologists, suitably represented as a graph, does not have a certain degree of connectivity, this implies that some biological data has been missed. (The converse is not true: even if the required connectivity does hold, there might of course be more information missing.)

Our result about k-connectedness being a useful test of missing information \cite{shukla} was obtained using SAT solvers for graphs up to a certain size, and a certain number of molecules. This was due to limitations in how we encoded the problem. Here we present a much more natural encoding that allows our result to be extended to graph sizes and molecule numbers that are typical of those found in real cells.

\label{subsec:graphmodel}
Since a VTS is a transport graph, it is but natural to formally model
VTS as graphs (as used in computer science) with their nodes denoting
compartments and labeled edges denoting transport vesicles with labels
denoting the set of molecules being transported. The pairing mechanism
can be represented as matching tables over sets of molecules.
%
Given such a graph model of VTSs and their properties, such as
stability condition and fusion rules, can be formally defined as
constraints over graphs and uninterpreted Boolean functions.
%
\

A VTS is {\em $k$-connected} if every pair of compartments remain
reachable after dropping $k-1$ vesicles.
This property of VTSs has been considered informative and
studied by~\cite{shukla}.
Here we have build an {\em efficient} tool that studies properties of
VTSs that are not $k$-connected, from some $k$. 

\input{pm-1-matrix}

\begin{example}
In figure~\ref{fig:M1}, we present a VTS that has 3 compartments and 8 molecules, and a corresponding pairing matrix.
Molecules are numbered 0-7.
In the VTS, labels are a string of molecule ids, and an overline over an id indicates that the molecule is active.
Every molecule on the node is active.
The activity of the molecules on an edge are controlled
by presence of the other molecules on the edge.
On the right side of the figure, we show the pairing matrix.
An entry 1 represents pairing between molecules.
$\times$ represents no pairing.
Rows corresponds to the labels of edges, and
columns corresponds to the labels of nodes.
Every molecule flows on a cycle, ensuring steady state.
This is a 3-connected graph.
\end{example}


%% file: pm-1-matrix.tex
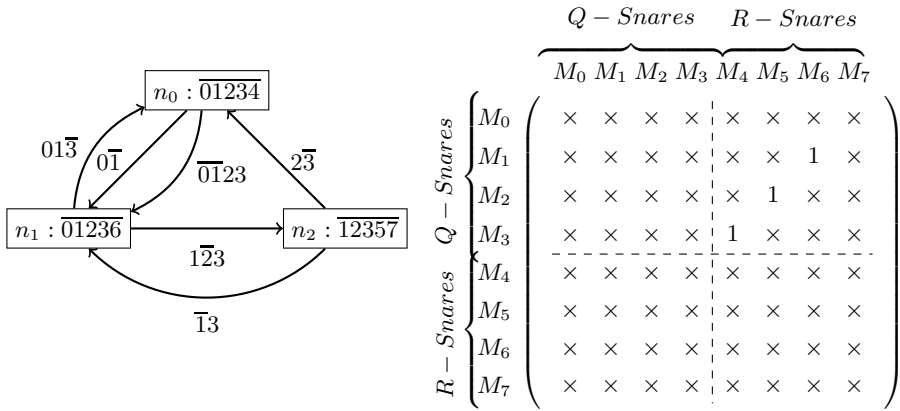
\begin{figure}[t]
\centering
\begin{minipage}{0.45\linewidth}

\begin{tikzpicture}[node distance = 26mm]
  \node[draw] (n0) {$n_0:\overline{01234}$};
  \node[draw,below left of=n0] (n1) {$n_1:\overline{01236}$};
  \node[draw,below right of=n0] (n2) {$n_2:\overline{12357}$};

  \draw[->,thick] (n0) -- node[left=1mm]  {$0\overline{1}$} (n1);
  \draw [->,thick] (n0) to [bend right=-30] node[right=1mm]  {$\overline{01}23$} (n1);
  \draw [->,thick] (n1) to [bend right=-30] node[left=1mm]  {$01\overline{3}$} (n0);

  \draw [->,thick] (n1) -- node[below=1mm]  {$1\overline{2}3$} (n2);
  \draw [->,thick] (n2) to [bend right=-45] node[below=2pt]  {$\overline{1}3$} (n1);

  \draw [->,thick] (n2) -- node[right=2pt]  {$2\overline{3}$} (n0);

\end{tikzpicture}
\end{minipage}
\begin{minipage}{0.50\linewidth}
  \center
\begin{tikzpicture}
\matrix [matrix of math nodes,left delimiter=(,right delimiter=),
         row sep=0.07cm,column sep=0.07cm] (m) {
\times & \times & \times &\times&\times&\times & \times & \times\\
\times & \times & \times&\times&\times&\times& 1 &\times \\
\times & \times & \times&\times&\times&1&\times &\times\\
\times & \times & \times&\times & 1 & \times &\times & \times\\
 \times & \times & \times & \times & \times & \times & \times & \times\\
 \times & \times & \times &\times & \times & \times & \times & \times\\
 \times & \times & \times & \times & \times & \times & \times & \times\\
 \times & \times & \times & \times & \times & \times&\times & \times\\};

\draw[dashed] ($0.5*(m-1-4.north east)+0.5*(m-1-5.north west)$) --
     ($0.5*(m-8-5.south east)+0.5*(m-8-4.south west)$);

\draw[dashed] ($0.5*(m-4-1.south west)+0.5*(m-5-1.north west)$) --
 ($0.5*(m-4-8.south east)+0.5*(m-5-8.north east)$);

\node[above=4pt of m-1-1] (top-1) {$M_0$};
\node[above=4pt of m-1-2] (top-2) {$M_1$};
\node[above=4pt of m-1-3] (top-3) {$M_2$};
\node[above=4pt of m-1-4] (top-4) {$M_3$};
\node[above=4pt of m-1-5] (top-5) {$M_4$};
\node[above=4pt of m-1-6] (top-6) {$M_5$};
\node[above=4pt of m-1-7] (top-7) {$M_6$};
\node[above=4pt of m-1-8] (top-8) {$M_7$};

\node[left=12pt of m-1-1] (left-1) {$M_0$};
\node[left=12pt of m-2-1] (left-2) {$M_1$};
\node[left=12pt of m-3-1] (left-3) {$M_2$};
\node[left=12pt of m-4-1] (left-4) {$M_3$};
\node[left=12pt of m-5-1] (left-5) {$M_4$};
\node[left=12pt of m-6-1] (left-6) {$M_5$};
\node[left=12pt of m-7-1] (left-7) {$M_6$};
\node[left=12pt of m-8-1] (left-8) {$M_7$};

\node[rectangle,above delimiter=\{] (del-top-1) at ($0.5*(top-1.south) +0.5*(top-4.south)$) {\tikz{\path (top-1.south west) rectangle (top-4.north east);}};
\node[above=10pt] at (del-top-1.north) {$Q-Snares$};
\node[rectangle,above delimiter=\{] (del-top-2) at ($0.5*(top-5.south) +0.5*(top-8.south)$) {\tikz{\path (top-4.south west) rectangle (top-6.north east);}};
\node[above=10pt] at (del-top-2.north) {$R-Snares$};

\node[rectangle,left delimiter=\{] (del-left-1) at ($0.5*(left-1.east) +0.5*(left-4.east)$) {\tikz{\path (left-1.north east) rectangle (left-4.south west);}};
\node[left=15pt,rotate=90,xshift=9mm] at (del-left-1.west) {$Q-Snares$};
\node[rectangle,left delimiter=\{] (del-left-2) at ($0.5*(left-5.east) +0.5*(left-8.east)$) {\tikz{\path (left-5.north east) rectangle (left-8.south west);}};
\node[left=15pt,rotate=90,xshift=9mm] at (del-left-2.west) {$R-Snares$};


\end{tikzpicture}  
\end{minipage}

\caption{An example of VTS and corresponding pairing matrix.} \label{fig:M1}
\end{figure}


%% file: encoding.tex
In this section, we will formally define VTSs as a labeled graph
and the discussed conditions on the graphs.
Subsequently, we will present the SMT encoding of the search problems
for the graphs that satisfy the conditions.

\subsection{Model}

We model VTSs as labeled directed graphs.
The graph labels both nodes and edges with sets of molecules to denote
the set of molecules present in them.
The graph is formally defined as follows.

\begin{df}
  A VTS $G$ is a tuple $(\nodes,\mols,\edges,\nlabel,\pairs,\edgef,\nodef)$, where
  \begin{itemize}
  \item $\nodes$ is a set of nodes representing compartments in the VTS,
  \item $\mols$ is the set of molecules flowing in the system, 
  \item $\edges \subseteq \nodes \times (\powerset{\mols}-\emptyset) \times \nodes$ is the
    set of edges with molecule sets as labels,
  \item $\nlabel : \nodes \maps \powerset{\mols}$ defines the molecules present in the nodes,
  \item $\pairs \subseteq \mols \times \mols$ is pairing relation,
  \item $\nodef : \mols \maps \powerset{\mols} \maps \booleans$ is activity maps for nodes, and
  \item $\edgef : \mols \maps \powerset{\mols} \maps \booleans $ is activity maps for edges.
  \end{itemize}
\end{df}
$\pairs$ defines which molecules can fuse with which molecules.
Let $\pairs(M')$ denote $\{m|(m,m') \in P \text{ and } m' \in M'\}$.
$\nodef$ and $\edgef$ are used to define activity of molecules on
nodes and edges respectively.
A molecule $k$ is {\em active} at node $n$ if $k \in \nlabel(n)$ and
$\nodef(k,\nlabel(n))$ is true.
A molecule $k$ is {\em active} at edges $(n,M',n')$ if $k \in M'$ and
$\edgef(k,M')$ is true.
We call $G$ {\em well-structured} if molecules $M$ is divided into
two equal-sized partitions $M_1$ and $M_2$ such that
$P \subseteq M_1 \times M_2$, and
for each $(n,M',n') \in \edges$, $n \neq n'$, 
$M' \subseteq \nlabel(n) \intersection \nlabel(n')$.
In other words, a well-structured VTS has no self loops, and 
each edge carry only those molecules that are present in its source
and destination nodes. 

We will also consider several variations of the model.
For example, unique edge between two nodes, activity of molecules is
not constrained by $\nodef$ and $\edgef$, etc.

A {\em path} in $G$ is a sequence $n_1,...,n_\ell$ of nodes 
such that $(n_i,\_,n_{i+1}) \in \edges$ for each $ 0 < i < \ell$.
For a molecule $m \in M$,
an {\em $m$-path} in $G$ is a sequence $n_1,...,n_\ell$ of nodes 
such that $(n_i,M',n_{i+1}) \in \edges$ and $m \in M'$ for
each $ 0 < i < \ell$.
A node $n'$ is {\em ($m$-)reachable} from node $n$ in $G$ if there is a ($m$-)path
$n,...,n'$ in $G$.
%
%
We call $G$ {\em stable} if for each $(n,M',n') \in \edges$ and $m \in M'$,
$n$ is $m$-reachable from $n'$.
An edge $(n,M',n') \in \edges$ {\em fuses} with its destination node $n'$
if there are molecules $m,m' \in \mols$ such that $m$ is active in
$(n,M',n')$, $m'$ is active in $n'$, and $(m,m') \in \pairs$.
We call $G$ {\em well-fused} if each edge $(n,M',n') \in \edges$ fuses
with non-empty fusing molecules $M'' \subseteq M'$
and $\pairs(M'')$ are not active in any other node.
We call $G$ {\em connected} if for each $n,n' \in \nodes$,
$n'$ is reachable from $n$ in $G$.
We call $G$ $k$-connected if for each $\edges' \subseteq \edges$ and $|\edges'| < k$,
VTS $(\nodes,\mols,\edges-E',\nlabel,\pairs,\edgef,\nodef)$ is connected.
In the definition, we do not care about the paths to be $m$-connected for some $m$.  
A variant of the definition may be sensitive to the $m$-connectedness, but
we are not considering the variation.

\paragraph{Search problem}
Based upon earlier discussion, we need to answer the following search
question among VTSs.
For a given $k$, size $\nu$, and molecule number $\mu$,
we are searching for well-structured, stable, and well-fused VTS
$G = (\nodes,\mols,\edges,\nlabel,\pairs,\edgef,\nodef)$ such that
$|\nodes| = \nu$, $|\mols| = \mu$, and
$G$ is not $k$-connected.

\subsection{Boolean satisfiability of the search problem}

We translate the search problem into a Boolean satisfiability
problem and use SMT solvers to find the satisfying VTSs.
We will first present the variables used to encode the
VTSs and the properties.
Afterwards, we will present the constraints corresponding to the
properties.

\subsubsection{Variables for VTS description}
We assume that size of the graph is $\nu$ and number of molecules is
$\mu$.
Furthermore, we also limit the maximum number $\pi$ of edges present
between two nodes.
Here, we list the vector of Boolean variables and uninterpreted function symbols
that encode the VTSs.
\begin{enumerate}

\item Boolean variable $n_{i,m}$ indicates if $m \in \nlabel(i)$
\item Boolean variable $e_{i,j,q}$ indicates if $q$th edge exists between $i$ and $j$.
\item Boolean variable $e_{i,j,q,m}$ indicates if $q$th edge between $i$ and $j$ contains $m$.
\item Boolean variable $p_{m,m'}$ indicates if $(m,m') \in \pairs$
\item uninterpreted Boolean functions $\nodef_m : \booleans^\mu \maps \booleans$
encoding $\nodef(m)$ map
\item uninterpreted Boolean functions $\edgef_m : \booleans^\mu \maps \booleans$
encoding $\edgef(m)$ map
\end{enumerate}
We also have auxiliary Boolean variables that will help us encode the well-fused and stability properties 
\begin{enumerate}
\item $a_{i,m}$ indicates that molecule $m$ is active at node $i$, i.e., $\nodef(m,L(i))$
  holds
\item $b_{i,j,q,m}$ indicates that molecule $m$ is active at $q$th edge $(i,M',j)$ between $i$ and $j$, i.e., $\edgef(m,M')$ holds
\item $r_{i,j,m,p}$ indicates if there is an $m$-path from
  $i$ to $j$ of length less than or equal to $p$
\end{enumerate}
For $k$-connected property, we also use the following auxiliary Boolean variables
\begin{enumerate}
\item $d_{i,j,q}$ indicates $q$th edge between $i$ and $j$ is dropped
\item $r'_{i,j}$ indicates if there is a path from $i$ to $j$ in the modified VTS
\end{enumerate}

We will describe the Boolean constraints that encode VTSs in several categories.
In the end, we will present in a table the constraints needed for the
model variants.
To avoid cumbersome notation, we will not explicitly write the ranges of the indexing
in the constraints.
$i$ and $j$ will range over nodes.
$m$ will range over molecules.
$q$ will range over edges between two nodes, i.e., from $1$ to $\pi$.

\subsubsection{Constraints on presence, activity of the molecule, and pairing matrix}
We need the following constraints~\eqref{eq:f0} and~\eqref{eq:c3}
to encode the basic structure of VTSs.
For an edge to exist it should have one molecule present. 
\begin{align}
  \bigwedge\limits_{i,j,q} (\bigvee_m e_{i,j,q,m} )\limplies e_{i,j,q}\tag{V1}\label{eq:f0}
\end{align}
If a molecule is active on an edge, it should be present on the edge.
\begin{align}
  \bigwedge\limits_{i,j,q,m} b_{i,j,q,m} \limplies e_{i,j,q,m}\tag{V2}\label{eq:f1}
\end{align}
A molecule should be present to be active on a node.  
\begin{align}
  \bigwedge\limits_{i,m} a_{i,m} \limplies n_{i,m}
  \tag{V3}\label{eq:c4}
\end{align}
The edge labels are subset of the node labels of the source and target nodes.
\begin{align}
  \bigwedge\limits_{i,j,q,m} e_{i,j,q,m} \limplies (n_{i,m} \land n_{j,m} )\tag{V4}\label{eq:c0}
\end{align}
Self edges are not allowed. 
\begin{align}
   \bigwedge\limits_{i,q} \neg e_{i,i,q}\tag{V5}\label{eq:c2}
\end{align}
We fix first half as Q-SNAREs and rest as R-SNAREs and set diagonal blocks in pairing matrix to be 0's.
\begin{align}
  \bigwedge\limits_{(x < M/2 \, \land  \, y < M/2) \lor  (x >= M/2 \land y >= M/2)} \neg p(x,y)
  \tag{V6}\label{eq:c3}
\end{align}

\subsubsection{Well-fused constraints}
Constraint~\eqref{eq:fuse1} encodes that each edge must fuse with
its destination node.
Constraint~\eqref{eq:fuse2} encodes that each edge should not
be able to potentially fuse with any node other than its destination node.
\begin{align}
  \bigwedge\limits_{i,j,q} e_{i,j,q} \limplies \bigvee_{m,m^{\prime}} (b_{i,j,q,m} \land a_{j,m^{\prime}} \land p_{m,m^{\prime}})
  \tag{V7}\label{eq:fuse1}  \\
\bigwedge\limits_{i,j,q,m} b_{i,j,q,m} \limplies \neg \bigvee_{j \neq j^{\prime}, m^{\prime\prime}} ( a_{j^{\prime},m^{\prime\prime}} \land p_{m,m^{\prime\prime}})
  \tag{V8}\label{eq:fuse2}  
\end{align}




\subsubsection{Regulation on nodes and edges}
We are considering several models that differ in constraints on
the activity of molecules.
We will present~\eqref{eq:ann}-~\eqref{eq:aep} that encodes
the varying constraints.
The following constraint encodes no conditions on activities on nodes,
i.e., all the present molecules on the nodes are active.
\begin{align}
\bigwedge\limits_{i,m} n_{i,m} = a_{i,m}    \tag{Ann}\label{eq:ann}
\end{align}
The following constraint encodes that activity of a molecule $m$ on the node is
defined by a Boolean function $\nodef_m$ of presence of molecules present on that node.
\begin{align}
\bigwedge\limits_{i,m} n_{i,m} \limplies a_{i,m} =  \nodef_m (n_{i,1},\dots,n_{i,\mu}) 
\tag{Anb}\label{eq:anb}
\end{align}
The following constraint encodes no conditions on activities on edges,
i.e., all the present molecules on the edges are active.
\begin{align}
  \bigwedge\limits_{i,j,q,m} e_{i,j,q,m} = b_{i,j,q,m}
\tag{Aen}\label{eq:aen}
\end{align}
The following constraint encodes that activity of a molecule $m$ on the edge is
defined by a Boolean function $\edgef_m$ of presence of molecules present on that edge.
\begin{align}
   \bigwedge\limits_{i,j,q,m} e_{i,j,q,m} \limplies b_{i,j,q,m} = \edgef_k(e_{i,j,q,1}, .., e_{i,j,q,\mu} )
  \tag{Aeb}\label{eq:aeb}
\end{align}
The following constraint encodes that the activity of the molecules on
edges is defined by inhibition by other molecules based on the pairing
matrix. 
\begin{align}
   \bigwedge\limits_{i,j,q, m}  [e_{i,j,q,m} \limplies  \bigwedge_{m' \neq m} (p_{m,m'} \limplies e_{i,j,q,m'})] \lequiv (\neg b_{i,j,q,m} \land\;\;\;\;  \bigwedge_{\mathclap{m' \neq m, p_{m,m'}}} \neg b_{i,j,q,m'})
  \tag{Aep}\label{eq:aep}
\end{align}
\subsubsection{Constraints for stability condition}
We use $m$-reachability to encode the stability condition in VTSs.
The following constraint recursively encodes that node $j$ is $m$-reachable from node $i$ in less than $p$ steps
if either there is a direct edge between $i$ and $j$ with $m$ present on the edge or there is a edge between $i^{\prime \prime}$ and
$(i \neq i^{\prime \prime})$ with $m$ present, and j is $m$-reachable from $i'$ in less than $p-1$ steps.
\begin{align}
  \bigwedge\limits_{i,j,m,p} r_{i,j,m,p} \limplies (\bigvee_{q} \, e_{i,j,q,m} \lor \bigvee_{i\neq i^{\prime}} ( \, \bigvee_{q} e_{i,i^{\prime},q,m}) \land r_{i^{\prime},j,m,p-1} )
  \tag{R1}\label{eq:reach1}
\end{align}
Now we can encode stability using the reachability variables.
and say if there is an $m$-edge between $i$ and $j$, there is
$m$-reachable path from $j$ to $i$.
\begin{align}
 \bigwedge\limits_{i,j,m} (\bigvee_{q} e_{i,j,q,m}) \limplies r_{j,i,m,\nu}
  \tag{R2}\label{eq:reach2}
\end{align}





\subsubsection{$k$-connectivity constraints}
To check whether $k$-connected is a necessary condition, we remove (drop) $k-1$ edges from the graph and if it
disconnects the graph, and we get an assignment. We have a graph that is not  $k$-connected.
Constraints~\ref{eq:drop1}-\ref{eq:drop4}
encode the relevant constraints for reachability
in the modified VTS. 
The following constraints encode that only
existing edges can be dropped and exactly $k-1$ edges are dropped.
\begin{align}
  \bigwedge\limits_{i,j,q} d_{i,j,q} \limplies e_{i,j,q}  \tag{D1}\label{eq:drop1}\\
  \sum_{i,j,q} d_{i,j,q} = k-1
  \tag{D2}\label{eq:drop2}
\end{align}
We need to define reachability in the modified VTS, therefore we use
a new variable $r'_{i,j}$ to encode reachability from $i$ to $j$.
In the following constraint, we encode $r'_{i,j}$ is true if there is an
edge $(i,\_,j)$ and it is not dropped, or there is a node
$i^{\prime}$ such that, there is an edge $(i,\_,i^{\prime})$ which is
not dropped and $r'_{i',j}$ is true.
\begin{align}
\bigwedge\limits_{i,j}  [\bigvee_{q} (e_{i,j,q} \land  \neg d_{i,j,q}) \lor  (\bigvee_{i' \neq i}  r^{\prime}_{i',j} \land  \bigvee_{q} (e_{i,i',q} \land \neg d_{i,i',q}) ] \limplies r^{\prime}_{i,j}  
  \tag{D3}\label{eq:drop3}
\end{align}
Since we search for disconnected modified VTS,
the following constraint encodes that there are 
unreachable pair of nodes in the underlying undirected graph.
\begin{align}
   \bigvee\limits_{i,j} \neg (r^{\prime}_{i,j} \lor r^{\prime}_{j,i})
  \tag{D4}\label{eq:drop4}
\end{align}

The key improvement in our work over the earlier tool is the
encoding of reachability, which was done using enumeration of paths.
In the current work, we have encoded reachability in two different
ways in constraints~\eqref{eq:reach1} and~\eqref{eq:drop3}.
The reachability is recursively defined in~\eqref{eq:drop3} and has
trivial solutions by making all $r'$s true.
However, the trivial solutions are disallowed by constraint~\eqref{eq:drop4} and we find
only the solutions that captures the evidence of unreachability.
On the other hand,
we have added length of paths in our reachability encoding in constraint~\eqref{eq:reach1},
which needs relatively more auxiliary variables.
This is because constraint~\eqref{eq:reach2} has only positive
occurrences of the reachability variables and if we had defined
$r$s in ~\eqref{eq:reach1} without paths,
the circular dependencies in the recursive definitions of $r$s
may have resulted in spurious satisfying assignments that
do not encode reachability.
By adding the path length, we break the circular dependencies, the
constraint remains polynomial in size, and satisfying assignments only
correspond to correct reachability.



\subsection{SMT problems for different variants}

We have encoded the following variants of VTSs.
The variants are due to different combinations of constraints on the
activity of molecules on the nodes and edges.
\begin{enumerate}[label=\Alph*]
\item Every present molecule is considered to be active.
\item Activity of molecules on the nodes is based on Boolean function of presence of other molecules. 
\item Activity of molecules on the edges is based on Boolean function of presence of other molecules
\item Activity of molecules both on the edges and nodes is based on Boolean function of presence of other molecules.
\item Activity of molecules on the edges is driven by pairing inhibition.
\item Activity of molecules on the nodes is based on Boolean function of presence of other molecules and on edge by pairing inhibition.
\end{enumerate}
%
%
In the table~\ref{tab:grph}, we present the constraints involved in each version.
The column two of the table shows that the constraints V1-V8, R1-R2, and D1-D4 are present
for every variant.
The column three of the table lists the constraints that are different among the variants.
One of the restriction is where the activity of the present molecule is dependent on the presence of the other molecules.
For example in version B, D, F activity of the molecules on the node is a boolean function of the presence of other molecules on that node; Anb.
Similarly for the case of the edge in version C and D; Aeb.
In the case of versions F the activity of the molecules on the edges
is described by pairing matrix; Aep.

%
The constraints for the variants can be given to a SMT solver to find
VTSs that belong to the variants.

%

%
%






%% file: experiments.tex
\begin{table}[t]
\centering
  \begin{tabular}{|c|c|c|c|}
    \hline
    {\multirow{2}{*}{\textbf{Variant}}}  & 
    \multicolumn{2}{c|}{\textbf{Constraints}} & 
    {\multirow{2}{*}{\textbf{Graph connectivity}}}
    \\
   \cline{2-3}
   &  \textbf{Rest} & \textbf{Activity} &  
\\ \hline
    
A. & \multirow{5}{*}{
     \makecell{ \ref{eq:f0}--\ref{eq:fuse2},\\
     \ref{eq:reach1},\ref{eq:reach2},\\
     \ref{eq:drop1}--\ref{eq:drop4}}
     } 
     & \ref{eq:ann},\ref{eq:aen} & No graph     \\\cline{3-4}
B. & & \ref{eq:anb},\ref{eq:aen} & No graph     \\\cline{3-4}
C. & & \ref{eq:ann},\ref{eq:aeb} & 3-connected  \\\cline{3-4}
D. & & \ref{eq:anb},\ref{eq:aeb} & 2-connected  \\\cline{3-4}
E. & & \ref{eq:ann},\ref{eq:aep} & No graph     \\\cline{3-4}
F. & & \ref{eq:anb},\ref{eq:aep} & 4-connected  \\\hline

  \end{tabular}
\caption{{\bf Activity regulation of molecules vs. graph connectivity.}}
\label{tab:grph}
\end{table}





%

We have implemented the encodings for each variants
using the python interface of $\zthree$ in a tool (MAA).
Our tool allows the user to choose a model and the size of the network
besides other parameters like connectivity and number of parallel
edges.
It uses Z3 Python interface to build the needed constraints and
applies $\zthree$ solver on the constraints to find a model (
a satisfying assignment that respects the constraints).
This tool also translates the satisfying model found by $\zthree$ into
a VTS and presents a visual output to the user in form of annotated graph.
%
%
%
We also visually report the dropped edges required to disconnect the graph, it gives information about the connectivity of the graph.


To illustrate usability of our tool 
in the last column of the table~\ref{tab:grph}, we present the minimum connectedness needed for
the different variants after applying our tool for sizes from 2 to 10. 
We found no graph for the variant A with constraints Ann and Aen.
Replacing constraint on the node of Ann with Anb (variant B) does not affect the outcome.
If we allow every present molecule to stay active (Ann) but constraint the edge
by a boolean function (Aeb) the resultant VTS has to be at least 3-connected. Similarly, the results for the other cases are presented in the table. 
%
%


\input{tbl-qf-graph}

In Table~\ref{tab:qf-grabh}, we present the running times for the search of
VTSs of sizes 2 to 10 that satisfy the variants.
and compare with our old encoding
({Old-e}) from~\cite{shukla}.
For the comparison between both encodings we have fixed the total
number of molecules to be $|M| = 2|N|$ for $ |N|> 2$ and
$|M| = 2|N| + 1$ for $|N| = 2$.
For each variant, we fix maximum number of parallel
edges to 2.
In the table we have shown comparison for specific connectivity, for
example variant A is checked against any graph with connectivity 2,
variant B with connectivity 3 similarly for the rest of the Variants.

The experiments were done on a machine with Intel(R) Core(TM) i3-4030U
CPU @ 1.90GHz processor and 4GB RAM.
We have compared our performance with the performance of our earlier 
CBMC based implementation (old-encoding).
For example, the formula for variation F, the total number of
compartments ($|N|$) equals to 10, returns in 129.78 minutes (7786.8 secs)
with a SAT result.
In comparison, CBMC results in OUT OF MEMORY for $|N|$ greater than 5.
``!'' indicate that the constraints were unsatisfiable.
Using this encoding in comparison to the old one, not only we got efficiency improved for finding a SAT model but also did better in the case of refutation that no model exist (Table 2 Variant A timing comparison and Variant D with N =2). Hence with the use of this novel encoding, we are able to scale the system to a much larger compartmentalized system, especially to
eukaryotic cells with a total number of 10 compartments.
%
%
Furthermore, we experimented with limits of our tools and found
that $\zthree$ was able to solve the constraints up to $\sim{14-18}$ nodes.







%% file: tbl-qf-graph.tex
\begin{table}[t]
  \centering
  \begin{tabular}[t]{|c|c|c|c|c|c|c|c|c|}\hline
  
    {\multirow{2}{*} {\textbf{Size}}}  & \multicolumn{2}{c|}{\textbf{Variant A}} & \multicolumn{2}{c|}{\textbf{Variant C}} & \multicolumn{2}{c|}{\textbf{Variant D}}  &  \multicolumn{2}{c|}{\textbf{Variant F}} \\\cline{2-9}
  
   {} & \multicolumn{2}{c|}{\textbf{2-connected} } & \multicolumn{2}{c|}{\textbf{3-connected} } & \multicolumn{2}{c|}{\textbf{2-connected}}  &  \multicolumn{2}{c|}{\textbf{4-connected}}
   \\\cline{2-9}
    {} & {\textbf{MAA}} & {\textbf{Old-e}} & {\textbf{MAA}} & {\textbf{Old-e}} & {\textbf{MAA}} & {\textbf{Old-e}} & {\textbf{MAA}} & {\textbf{Old-e}} \\\hline
    2 & !0.085 & !2.43 & 0.15 & 2.12 & !0.13 & !1.89 & 0.35 & 5.12 \\\hline
    3 & !0.54 & !8.04 & 0.95  & 7.65 & 0.62 & 7.66  & 1.36 & 23.94\\\hline
    4 & !2.57 & !297.93 & 2.33 & 22.74 & 2.85 & 48.35  & 4.81 & 123.34\\\hline
    5 & !7.7 & !3053.8 & 7.60 & 500.03 & 10.27 & 890.84 & 33.36  & 2482.71 \\\hline
    6 & !22.98 & M/O & 19.52 & M/O & 30.81 & M/O  & 147.52 & M/O\\\hline
    7 & !57.07 & M/O & 81.89 & M/O & 82.94 & M/O & 522.26  & M/O \\\hline
    8 & !164.14 & M/O & 630.85 & M/O & 303.19 & M/O & 2142.76 & M/O\\\hline
    9 & !307.67 & M/O & 2203.45 & M/O & 971.01 & M/O & 4243.34 & M/O\\\hline
    10 & !558.34 & M/O & 7681.93 & M/O & 2274.30 & M/O & 7786.82 & M/O\\\hline
  \end{tabular}
  \caption{Run-times for searching for models (in secs).}
  \label{tab:qf-grabh}
\end{table}

%% file: related.tex
Modern day SAT/SMT solvers can handle millions of variables and due to the exhaustive nature of the searching, solving the combinatorial problem is a natural fit for the SAT/SMT solvers. There are many graphs related combinatorial work \cite{gay2013solving,wotzlaw2012generalized} that have used SAT solvers. Thanks to the improvement in SAT solver and formal techniques, tools like model checkers and theorem provers scalability and combinatorial solving capability are now being used to model complex biological systems \cite{heule2010exact,yordanov2013smt,mangla2010timing}.
Many of these biological system uses these tools and techniques to reason about graphs and networks rules with possible exhaustive search \cite{guerra2012reasoning,chin2008biographe}.  

Recently these tools and techniques have found the use of modeling and understanding the gene regulatory networks (GNR) \cite{giacobbe2015model,rosenblueth2014inference, batt2010efficient, yordanov2016method, dunn2014defining, paoletti2014analyzing, koksal2013synthesis}. GNR is a complex system driven by many complex rules. We extended this work and have applied model checkers to the more complex transport network (VTS) \cite{mani2016stacking,shukla}, but scaling it to cases which are interesting biologically is still a challenge. In this paper, we have achieved the scalability required for analyzing VTS for eukaryotic cells, of total ten compartments (N = 10). \\



%% file: conclusion.tex
In the experiments, we have demonstrated the power of SMT solvers and the value
of careful encoding of the problems into SMT queries.
We manage to scale the tool upto the size that makes tool
biologically relevant.
However, there are many further search problems or extensions
that are of interest.
For example, are there any k-connected graphs that can not be
annotated into a VTS?
This problem induces a quantifier alternation in an encoding.
Therefore, a simple call to SAT solver will not work.
We are planning to use QBF(quantified Boolean formulas) solvers or develop
iterative search algorithm for such queries.

One may be interested in counting the number of graphs that satisfy
a given property.
Exact counting of the graphs using SAT solvers may prove to be very difficult.
We are also planning to employ some methods for approximate counting of solutions.





%% file: main.bbl
\begin{thebibliography}{10}

\bibitem{stenmark2009rab}
Harald Stenmark.
\newblock Rab gtpases as coordinators of vesicle traffic.
\newblock {\em Nature reviews Molecular cell biology}, 10(8):513--525, 2009.

\bibitem{dacks2007evolution}
Joel~B Dacks and Mark~C Field.
\newblock Evolution of the eukaryotic membrane-trafficking system: origin,
  tempo and mode.
\newblock {\em Journal of cell science}, 120(17):2977--2985, 2007.

\bibitem{munro2004organelle}
Sean Munro.
\newblock Organelle identity and the organization of membrane traffic.
\newblock {\em Nature cell biology}, 6(6):469--472, 2004.

\bibitem{mani2016stacking}
Somya Mani and Mukund Thattai.
\newblock Stacking the odds for golgi cisternal maturation.
\newblock {\em Elife}, 5:e16231, 2016.

\bibitem{shukla}
Ankit Shukla, Arnab Bhattacharyya, Lakshmanan Kuppusamy, Mandayam Srivas, and
  Mukund Thattai.
\newblock Discovering vesicle traffic network constraints by model checking.
\newblock {\em PloS one}, 12(7):e0180692, 2017.

\bibitem{clarke1996symbolic}
E~Clarke, K~McMillan, S~Campos, and Vasiliki Hartonas-Garmhausen.
\newblock Symbolic model checking.
\newblock In {\em Computer Aided Verification}, pages 419--422. Springer, 1996.

\bibitem{biere2003bounded}
Armin Biere, Alessandro Cimatti, Edmund~M Clarke, Ofer Strichman, and Yunshan
  Zhu.
\newblock Bounded model checking.
\newblock {\em Advances in computers}, 58:117--148, 2003.

\bibitem{clarke2008birth}
Edmund Clarke.
\newblock The birth of model checking.
\newblock {\em 25 Years of Model Checking}, pages 1--26, 2008.

\bibitem{cimatti2000nusmv}
Alessandro Cimatti, Edmund Clarke, Fausto Giunchiglia, and Marco Roveri.
\newblock Nusmv: a new symbolic model checker.
\newblock {\em STTT}, 2(4):410--425, 2000.

\bibitem{holzmann1997model}
Gerard~J. Holzmann.
\newblock The model checker spin.
\newblock {\em IEEE Transactions on software engineering}, 23(5):279--295,
  1997.

\bibitem{moskewicz2001chaff}
Matthew~W Moskewicz, Conor~F Madigan, Ying Zhao, Lintao Zhang, and Sharad
  Malik.
\newblock Chaff: Engineering an efficient sat solver.
\newblock In {\em DAC}. ACM, 2001.

\bibitem{een2004extensible}
Niklas E{\'e}n and Niklas Sorensson.
\newblock An extensible sat-solver.
\newblock {\em Lecture notes in computer science}, 2919:502--518, 2004.

\bibitem{CKY03}
Daniel Kroening, Edmund Clarke, and Karen Yorav.
\newblock Behavioral consistency of {C} and {Verilog} programs using bounded
  model checking.
\newblock In {\em DAC}. ACM, 2003.

\bibitem{ckl2004}
Edmund Clarke, Daniel Kroening, and Flavio Lerda.
\newblock A tool for checking {ANSI-C} programs.
\newblock In {\em TACAS}, pages 168--176. Springer, 2004.

\bibitem{z3}
Leonardo de~Moura and Nikolaj Bjorner.
\newblock Z3: An efficient smt solver.
\newblock In {\em TACAS}, volume 4963 of {\em LNCS}, pages 337--340. Springer
  Berlin Heidelberg, 2008.

\bibitem{gay2013solving}
Steven Gay, Fran{\c{c}}ois Fages, Francesco Santini, and Sylvain Soliman.
\newblock Solving subgraph epimorphism problems using clp and sat.
\newblock In {\em WCB-ninth Workshop on Constraint Based Methods for
  Bioinformatics, colocated with CP}, pages 67--74, 2013.

\bibitem{wotzlaw2012generalized}
Andreas Wotzlaw, Ewald Speckenmeyer, and Stefan Porschen.
\newblock Generalized k-ary tanglegrams on level graphs: A satisfiability-based
  approach and its evaluation.
\newblock {\em Discrete Applied Mathematics}, 160(16):2349--2363, 2012.

\bibitem{heule2010exact}
Marijn Heule and Sicco Verwer.
\newblock Exact dfa identification using sat solvers.
\newblock {\em Grammatical Inference: Theoretical Results and Applications},
  pages 66--79, 2010.

\bibitem{yordanov2013smt}
Boyan Yordanov, Christoph~M Wintersteiger, Youssef Hamadi, and Hillel Kugler.
\newblock Smt-based analysis of biological computation.
\newblock In {\em NASA formal methods symposium}, pages 78--92. Springer, 2013.

\bibitem{mangla2010timing}
Karan Mangla, David~L Dill, and Mark~A Horowitz.
\newblock Timing robustness in the budding and fission yeast cell cycles.
\newblock {\em PLoS One}, 5(2):e8906, 2010.

\bibitem{guerra2012reasoning}
Jo{\~a}o Guerra and In{\^e}s Lynce.
\newblock Reasoning over biological networks using maximum satisfiability.
\newblock In {\em PPCP}, pages 941--956. Springer, 2012.

\bibitem{chin2008biographe}
George Chin, Daniel~G Chavarria, Grant~C Nakamura, and Heidi~J Sofia.
\newblock Biographe: high-performance bionetwork analysis using the biological
  graph environment.
\newblock {\em BMC bioinformatics}, 9(6):S6, 2008.

\bibitem{giacobbe2015model}
Mirco Giacobbe, Calin~C Guet, Ashutosh Gupta, Thomas~A Henzinger, Tiago Paixao,
  and Tatjana Petrov.
\newblock Model checking gene regulatory networks.
\newblock In {\em TACAS}, 2015.

\bibitem{rosenblueth2014inference}
David~A Rosenblueth, Stalin Mu{\~n}oz, Miguel Carrillo, and Eugenio Azpeitia.
\newblock Inference of boolean networks from gene interaction graphs using a
  sat solver.
\newblock In {\em ICACB}, pages 235--246. Springer, 2014.

\bibitem{batt2010efficient}
Gregory Batt, Michel Page, Irene Cantone, Gregor Goessler, Pedro Monteiro, and
  Hidde De~Jong.
\newblock Efficient parameter search for qualitative models of regulatory
  networks using symbolic model checking.
\newblock {\em Bioinformatics}, 26(18):i603--i610, 2010.

\bibitem{yordanov2016method}
Boyan Yordanov, Sara-Jane Dunn, Hillel Kugler, Austin Smith, Graziano Martello,
  and Stephen Emmott.
\newblock A method to identify and analyze biological programs through
  automated reasoning.
\newblock {\em NPJ systems biology and applications}, 2:16010, 2016.

\bibitem{dunn2014defining}
S-J Dunn, Graziano Martello, Boyan Yordanov, Stephen Emmott, and AG~Smith.
\newblock Defining an essential transcription factor program for naive
  pluripotency.
\newblock {\em Science}, 344(6188):1156--1160, 2014.

\bibitem{paoletti2014analyzing}
Nicola Paoletti, Boyan Yordanov, Youssef Hamadi, Christoph~M Wintersteiger, and
  Hillel Kugler.
\newblock Analyzing and synthesizing genomic logic functions.
\newblock In {\em International Conference on Computer Aided Verification},
  pages 343--357. Springer, 2014.

\bibitem{koksal2013synthesis}
Ali~Sinan Koksal, Yewen Pu, Saurabh Srivastava, Rastislav Bodik, Jasmin Fisher,
  and Nir Piterman.
\newblock Synthesis of biological models from mutation experiments.
\newblock In {\em ACM SIGPLAN Notices}, volume~48, pages 469--482. ACM, 2013.

\end{thebibliography}
